\documentstyle[twocolumn,aps,floats,amssymb,epsfig]{revtex}

\newcommand{\beq}{\begin{equation}}
\newcommand{\eeq}{\end{equation}}
\newcommand{\beqa}{\begin{eqnarray}}
\newcommand{\eeqa}{\end{eqnarray}}

\newcommand{\avd}{\langle\tau\rangle}

\def\half{\mbox{$\frac12$}}
\def\binomial#1#2{\left({#1 \atop #2}\right)}
\def\eref#1{(\protect\ref{#1})}
\def\etal{{\it{}et~al.}}
\def\O{{\rm O}}

\newdimen\figurewidth
\begin{document}

\twocolumn[\hsize\textwidth\columnwidth\hsize\csname @twocolumnfalse\endcsname

\title{Glassiness and constrained dynamics of a short-range
non-disordered spin model}

\author{
Juan P. Garrahan$^1$ 
and
M. E. J. Newman$^2$
}

\address{
$^1$ Theoretical Physics, University of Oxford,
1 Keble Road, Oxford OX1 3NP, United Kingdom \\
$^2$ Santa Fe Institute, 1399 Hyde Park Road, Santa Fe, NM 87501, U.S.A.
}

\maketitle

\begin{abstract}
  We study the low temperature dynamics of a two dimensional
  short-range spin system with uniform ferromagnetic interactions, which
  displays glassiness at low temperatures despite the absence of disorder
  or frustration.  The model has a dual description in terms of free
  defects subject to dynamical constraints, and is an explicit realization
  of the ``hierarchically constrained dynamics'' scenario for glassy
  systems.  We give a number of exact results for the statics of the model,
  and study in detail the dynamical behaviour of one-time and two-time
  quantities.  We also consider the role played by the configurational
  entropy, which can be computed exactly, in the relation between
  fluctuations and response.
\end{abstract}

\pacs{}

]

\section{Introduction}
Understanding the nature of the low-temperature dynamics of glasses and
other strongly interacting many body systems remains one of the outstanding
open problems in condensed matter and statistical
physics~\cite{angell,review}.

Consider, for example, the case of glass-forming supercooled liquids.  One
of the reasons for glassiness in such systems is that the rearrangements of
atoms necessary for their relaxation involves activation over energy
barriers.  It is natural to plot the logarithm of the viscosity or
equilibration time against the inverse temperature, giving a so-called
Arrhenius plot, which should take a straight-line form if the barrier
heights remain constant with varying temperature.  For ``strong''
liquids~\cite{angell}, such as SiO$_2$, the Arrhenius plot is indeed a
straight line, but for ``fragile'' liquids it is not, following instead the
empirical Vogel--Fulcher law $\exp[\mbox{const.}/(T-T_0)]$, although other
forms not displaying a finite $T$ singularity can be fit too~\cite{temps}.
If the dynamics of fragile liquids is due to activation (which is not the
only possible explanation), this implies that the energy barriers grow with
decreasing temperature, presumably because they increase with the
increasing size of correlated regions in the system.

Another set of interesting questions, most relevant for fragile liquids,
revolves around the possible existence of an ideal (continuous) phase
transition to a true thermodynamic glass state at some temperature~$T_K$
(the Kauzmann temperature) lying below the glass
temperature~\cite{gibbs,jarev}.  Although the existence of such a
transition would resolve Kauzmann's paradox, in which the extrapolation of
the configurational entropy of the supercooled liquid appears to pass below
that of the crystal~\cite{kauzmann}, and is also supported by the analogy
between fragile glasses and discontinuous mean field
spin-glasses~\cite{ktw}, there is some evidence against it.  For example,
it has been shown numerically~\cite{krauth} that a thermodynamic phase
transition is absent for a system of polydisperse hard-disks, a typical
model glass former.  It has also been argued that the extrapolations which
yield a positive $T_K$ in fragile liquids composed of molecules of finite
size are flawed~\cite{stilli}.

Given the many interesting open questions in this field, it is important to
find simple and if possible solvable models for the various features
displayed by glasses.  The discontinuous mean-field spin-glasses, such as
the spherical $p$-spin model~\cite{pspin}, provide a good example of such a
model system.  However, while some dynamical features of these
models~\cite{review,cuku} are also observed in simulations of Lennard-Jones
glasses~\cite{kob,lapr}, the presence of disorder in the models and,
crucially, of long-range interactions limits their usefulness as models of
systems which are intrinsically short-ranged and disorder-free, like hard
spheres in two and three dimensions.

An alternative approach to modeling these latter systems is the
``hierarchically constrained dynamics'' of Palmer~\etal~\cite{palmer},
in which it is hypothesized that, for a strongly interacting system
displaying glassy behaviour, it should be possible to describe the
dynamics in terms of hierarchies of degrees of freedom, from fast to
slow, independent of the presence of disorder or even frustration.
These hierarchies would be weakly interacting in the energetic sense,
but their dynamics would be constrained, the faster modes constraining
the slower ones.  It is known that the presence of kinetic constraints
in the dynamics can directly induce glassiness.  A good example is the
kinetically constrained Ising chain~\cite{fa,jackle,sollich}, for
which the Hamiltonian is trivial, but the transition rates for the
flipping of individual Ising spins depend on the states of
neighbouring spins.  Kinetic constraints can also arise as a result of
entropic barriers, as in the Backgammon model, for
instance~\cite{backgammon}.  Fragile glass behaviour has also been
observed in kinetic lattice-gas models \cite{kps,sellito}.  Although
interesting in their own right, these models have somewhat {\it ad
hoc\/} dynamics, and represent only one side of the scenario discussed
in Ref.~\onlinecite{palmer}.  A recent comparative study of
one-dimensional constrained kinetic models has been given by
Crisanti~\etal~\cite{felix}.

In this paper, we study in detail a model introduced by Newman and
Moore~\cite{tri}, which is a form of two-dimensional Ising model with
uniform short-range ferromagnetic interactions.  Despite the absence of
either disorder or frustration, this model displays glassy behaviour at low
temperatures.  The cause of this behaviour is the presence of energy
barriers which grow logarithmically with the size of correlated regions.
The model has a dual description in terms either of strongly interacting
spins subject to simple single-spin-flip dynamics, or of free ``defects''
subject to a constrained dynamics.  It is thus an explicit realization of
the constrained dynamics scenario of Palmer~\etal~\cite{palmer}.  Other
non-disordered short-range spin systems displaying glassy features are
three-dimensional Ising models with competing nearest and next-nearest
neighbour interactions~\cite{shore}, or with ferromagnetic four-spin
plaquette interactions~\cite{lipo,lipoje,swift}.

The paper is organized as follows.  In Section~\ref{statics} we describe
the model and extend the exact solution of the statics given in
Ref.~\onlinecite{tri} by calculating all equilibrium spin correlation
functions, and showing that the phase-transition to an ordered state occurs
only at zero temperature.  In Section~\ref{dynamics}, we give our principal
results, which concern the out-of-equilibrium dynamics of the system
following a quench from a random configuration to low temperatures.  We
show that the equilibration time diverges with an exponential inverse
temperature squared law, similar to that found for the asymmetrically
constrained Ising chain.  We study the behaviour of one-time quantities,
like energy density and spin correlations and two-time quantities, like
autocorrelation and response functions.  We also study the
out-of-equilibrium fluctuation-dissipation relations and relate them to the
configurational entropy, which can be calculated exactly.  Our conclusions
are given in Section~\ref{concs}.

\section{Model and Static Solution}
\label{statics}
We consider the model introduced in Ref.~\onlinecite{tri}, which consists
of Ising spins $\sigma=\pm1$ on a triangular lattice with uniform
short-range three-spin ferromagnetic interactions: each spin interacts only
with its nearest-neighbours, and only in groups of three lying at the
vertices of a downward-pointing triangle on the lattice.  Note that this is
distinct from the model of Baxter and Wu~\cite{baxterwu}, which has
interactions on upward-pointing triangles also (and which is not glassy).
The Hamiltonian for the model is
\begin{equation}
H = \half J \, \sum_{mn} \sigma_{mn} \, \sigma_{m,n+1} \, \sigma_{m-1,n+1} 
        + \half NJ,
\end{equation} 
where the indices $m$ and $n$ run along the unit vectors of the lattice
$\vec{a}_1 \equiv \hat{x}$ and $\vec{a}_2 \equiv \frac{1}{2} (\hat{x} +
\sqrt{3} \hat{y})$.  The constant in the Hamiltonian is added for
convenience to make the minimum possible energy equal to zero.  The model
can also be formulated using the defect variables
\begin{equation}
\tau_{mn} \equiv \sigma_{mn} \, \sigma_{m,n+1} \, \sigma_{m-1,n+1},
\label{defects}
\end{equation}
in terms of which the Hamiltonian is
\begin{equation}
H = \half J \, \sum_{mn} \tau_{mn} + \half NJ.
\label{h2}
\end{equation} 
On lattices of size a power of two in at least one direction, with periodic
boundary conditions, there is a one-to-one correspondence between spin and
defect configurations, and hence the partition function is given by
\begin{equation}
Z = (2 \, e^{-\frac12\beta} \, \cosh\half\beta)^N,
\end{equation}
where $N$ is the number of spins, and we set $J=1$ from here on.  The
equilibrium energy density is then
\begin{equation}
\varepsilon_{\rm eq} 
        = \half \bigl( 1 +  \avd \bigr)
        = \half ( 1 - \tanh{\half\beta}) .
\label{eqen}    
\end{equation}

We now wish to invert Eq.~\eref{defects} and write the spins as functions
of the defects.  It is known~\cite{tri} that if the spins are represented
by the variables $s_{mn} \equiv \half (\sigma_{mn}+1) \in \{0,1\}$ the
spins below an isolated defect on the lattice form a Pascal's triangle
mod~2 (i.e.,~a triangular array with the binomial coefficients
$\binomial{n}{r}$ mod~2 as entries).  This means that the value of the spin
$s_{mn}$ is given by the superposition mod~2 of the Pascal's triangles of
all defects $\tau_{kl}=+1$ with $m-l\le k\le m$ and $l\ge n$.  The inverse
of the transformation~\eref{defects} then reads
\begin{equation}
\sigma_{mn} = - \!\!\!\prod_{{n\le l}\atop{m-l\le k\le m}}
                \!\!\!(-\tau_{kl})^{\binomial{l-n}{m-k}}.
\label{spin}
\end{equation}
With this result we can compute the equilibrium correlation functions of
the spins.  First, the magnetization is given by averaging~\eref{spin}:
\begin{equation}
\langle \sigma_{mn} \rangle = - (\tanh\half\beta)^{{\cal N}_{mn}}
\end{equation}
where
\begin{equation}
{\cal N}_{mn} = \sum_{{n\le l}\atop{m-l\le k\le m}}
                \mbox{$\binomial{l-n}{m-k}$ mod~2}
\end{equation}
is the total number of ones in the inverted Pascal's triangle with its tip
at site $(m,n)$.  This number diverges faster than the linear size of the
triangle, so in the thermodynamic limit we obtain
\begin{equation}
\langle \sigma_{mn} \rangle = 
        \left\{ 
        \begin{array}{cl} 
        -1 \qquad & \mbox{for $T=0$} \\
         0 \qquad & \mbox{for $T>0$,}
        \end{array} 
        \right.
\end{equation}
for all $(m,n)$, implying that the system has a $T=0$ static phase
transition.

Arbitrary correlation functions can be computed in a similar manner:
\begin{equation}
\langle \sigma_{mn} \cdots \sigma_{m'n'} \rangle 
        = - (\tanh\half\beta)^{{\cal N}_{mn \cdots m'n'}}
\end{equation}
where now ${\cal N}_{mn \cdots m'n'}$ is the total number of ones in the
superposition mod~2 of the inverted Pascal's triangles starting in
positions $(m,n)$ through $(m',n')$.  Since $\binomial{l}{0} =
\binomial{l}{l} = 1$, the left and right edges of a Pascal's triangle
contain only ones, so any superposition of two triangles always has an
infinite number of ones coming from the edges.  This implies that {\em
  all\/} 2-spin correlations vanish at $T>0$.  The first non-zero
correlations are those for three spins at the vertices of inverted
equilateral triangles of side $2^k$.  In this case ${\cal N} = 3^k$, and
the correlation is
\begin{equation}
C_k^{(3)} = \langle \sigma_{mn} \, \sigma_{m,n+2^k} \,
                    \sigma_{m-2^k,n+2^k} \rangle 
        = - (\tanh\half\beta)^{3^k}.
\label{c3}
\end{equation}
Notice that
\begin{equation} 
C_{k+1}^{(3)}(\tanh\half\beta)=C_k^{(3)}(\tanh^3\half\beta).
\label{c3k}
\end{equation}
A similar scaling relation is seen in the one-dimensional Ising model.

At low temperatures the system is in a scaling region, and using
Eq.~\eref{c3} we find a correlation length of
\begin{equation}
\xi = (\ln \coth\half\beta)^{-\ln2/\!\ln3}.
\label{xi}
\end{equation}

\section{Dynamics}
\label{dynamics}
Newman and Moore~\onlinecite{tri} found that the model studied here shows
glassy behaviour under a single-spin-flip dynamics.  They showed that
following a quench from $T=\infty$ the system was unable to equilibrate in
finite time at low enough temperature ($T\lesssim0.2$ for $t_{\rm max} \sim
10^9$ in their simulations).  The system also fell out of equilibrium for
exponential cooling with a variety of cooling rates.  The reason for this
glassiness is to be found in the details of the system's dynamics.
Single-spin flips correspond to flips of the defect variables on triples of
sites forming upward-pointing triangles on the lattice, and sets of such
flips can be combined to flip upward-pointing triangles of side $2^k$ for
any integer~$k$.  Any isolated such triangle with $k>0$ is locally stable;
in order to remove it we have to cross an energy barrier of height~$k$.
Thus as the system relaxes it has to cross energy barriers which grow
logarithmically with the size of the equilibrated regions.

Another way of looking at this is to observe that at low temperatures the
flipping of an isolated defect is heavily suppressed, since such a flip
requires the creation of two other defects and so incurs a net energy
penalty.  Thus, a defect requires the presence of another neighbouring
defect to be able to flip, a situation highly reminiscent of the
facilitated kinetic Ising model of Ref.~\onlinecite{fa}; the defects are
non-interacting in the Hamiltonian, but their low temperature dynamics is
effectively constrained.

At low temperatures excitations of linear size larger than~1 can only be
annihilated via activation.  From the observation that barriers grow
logarithmically with linear size it is straightforward to estimate
equilibration time.  The rate of relaxation of an excitation of linear size
$d$ is given by the Arrhenius formula $\Gamma(d) \sim \exp(-\ln d/
T\ln2)$.  Thus after time $t$ the average linear distance between defects
is $d \sim t^{T\ln2}$.  The equilibrium value of this distance at low $T$
is $d_{\rm eq} \sim \exp(\beta/2)$ [see Eq.~\eref{eqen}], and hence the
equilibration time is
\begin{equation}
t_{\rm eq} \sim \exp\left(\frac{1}{2\,T^2\,\ln 2}\right).
\label{teq}
\end{equation}
This exponential inverse temperature squared (EITS) form is similar to
that obtained in Ref.~\onlinecite{sollich} for the asymmetrically
constrained Ising chain (ACIC)~\cite{jackle}, except for the~2 in the
denominator which is due to the two-dimensional nature of our model.
The EITS form has no finite-temperature singularity, which is
consistent with the fact that our model has no finite-temperature
phase transition.  The behaviour of the relaxation, however, is
``fragile''.  In fact, the EITS form can be approximated by a
Vogel--Fulcher form with $T_0 \sim \frac23 T_g$.  A system which has
an analogous but ``strong'' behaviour is the Frenkel-Kontorova model
\cite{sethna}: at low temperatures above its $T=0$ phase transition,
the relaxation of defects is hindered by the presence of constant
energy barriers, and the equilibration time diverges in an Arrhenius
way.

Equation~\eref{teq} clarifies the observations made in
Ref.~\onlinecite{tri}.  In the case of relaxation following a quench
from $T_0=\infty$ to $T$, the highest temperature at which the system
was found to equilibrate within time $t_{\rm max} = 10^9$ was $T=0.2$,
while for $T = 0.18$ or smaller the system was unable to reach
equilibrium.  This is explained by the fact that $t_{\rm eq}(T=0.2)
\sim 10^8$, while $t_{\rm eq}(T=0.18) \sim 10^{10}$.  In the case of
exponential cooling, our expression for $t_{\rm eq}$ implies a maximum
allowable cooling rate of $\gamma_{\max}(T) \sim 1/t_{\rm eq}(T)$ if
we wish to remain in equilibrium at temperatures $T$ and greater.  For
the different annealing simulations of \cite{tri}, the temperatures at
which the energy ceases to follow the equilibrium curve are given by
the solution of $\gamma_{\rm max}(T)=\gamma$ for the cooling rate
$\gamma$ used.

\subsection{One time quantities}
\label{onetime}
We now consider the behaviour of one-time quantities for the model, such as
energy density and spin correlations, following a quench from $T=\infty$ to
a low temperature.

\begin{figure}[t]
\begin{center}
\epsfig{file=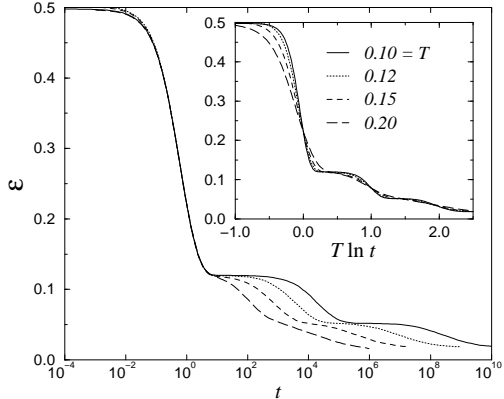,width=\figurewidth}
\end{center}
\caption{Energy density $\varepsilon=\langle H \rangle /N$ as a function
  of time following a quench from $T=\infty$ to $T=0.10$, $0.12$, $0.15$
  and $0.20$, from Monte Carlo simulations for a system of $256\times256$
  spins.  Time is in Monte Carlo steps per spin,
  and all magnitudes are in dimensionless units.  
  Inset: the same data as a
  function of rescaled time $\nu = T \ln t$.}
\label{et}
\end{figure}

The $T=\infty$ configuration is random, and therefore has zero correlation
length.  As the system relaxes following the quench it tries to increase
its correlation length to the equilibrium value, Eq.~\eref{xi}, which is
finite but large at low temperature.  In terms of the defects, the system
is trying to decrease the defect density or equivalently the internal
energy per spin $\varepsilon$ from its initial value of $\varepsilon_0 =
\half$.
In Fig.~\ref{et} we show numerical results for the evolution in time of the
energy density after a quench from $T=\infty$ to a variety of final
temperatures.  The simulations were performed on a $256\times256$ rhombic
lattice using a Bortz--Kalos--Lebowitz continuous time
algorithm~\cite{bkl}.  As the figure shows, after an initial
temperature-independent exponential relaxation corresponding to the
(barrierless) removal of pairs of neighbouring defects, the energy displays
``plateaus'', which are more pronounced the lower the temperature.  These
plateaus are the result of the system becoming trapped in locally stable
configurations.  As we can see, the time taken by the system to escape
these plateaus becomes larger with decreasing temperature.

As shown in the previous section, the typical length-scale in the system
(i.e.,~the typical distance separating defects) grows with time as $t^T$,
and in the absence of other length-scales we might thus expect the data in
Fig.~\ref{et} approximately to collapse when plotted against $t^T$, or
equivalently against $\nu=T\ln t$ on the logarithmic scales used in the
figure.  (A collapse of this kind was found for the ACIC in
Ref.~\onlinecite{sollich}.)  In the inset of Fig.~\ref{et} we show that
this is indeed the case for our model.  Save for the initial exponential
transient, the collapse of the curves for different temperatures is very
good.  Moreover, the first plateau seems to extend from $\nu=0$ to $1$, the
second from $\nu=1$ to~$2$, and so on (a behaviour also seen in the ACIC).

At low temperatures, the relaxation of the system can also be regarded as
an anomalous coarsening process.  For the case considered here of periodic
boundary conditions, there exists a unique minimum energy configuration of
the model in which all the spins point down and there are no defects.  For
free boundary conditions, however, there are $2^{2L-1}$ degenerate minima.
(The spins along the bottom and right hand side of our rhombic system, for
instance, may be chosen arbitrarily, with the rest being uniquely fixed by
the requirement that there be no defects.)  Following a quench, spins start
to rearrange locally to eliminate defects, regardless of the boundary
conditions, and thus form domains of the various free-boundary minima. The
imperfect matching of these domains will be marked by the presence of
defects.

The coarsening dynamics for $T$ close to zero can be studied approximately
using the generating function method of Sollich and Evans~\cite{sollich}.
Consider a description of a configuration of the system in terms of domains
containing no defects ($\tau=-1$) bounded by sites containing defects
($\tau=1$).  In contrast with the one-dimensional model studied
in~\onlinecite{sollich}, such a description cannot be made exact for our
model.  Nevertheless, as far as average properties go, such as distance
between defects, we can think of low energy configurations as an
approximate tiling of defect-free parallelograms each delimited by a defect
in, say, its top right-hand corner.  This effectively maps the problem into
one dimension.  As we now show, this crude approximation, which allows us
to apply the method of Ref.~\onlinecite{sollich}, gives reasonable results.

When $T\to0$, we have $\varepsilon_{\rm eq}\to0$ and the plateaus of
Fig.~\ref{et} become distinct stages in the dynamics.  During stage $n$,
all domains of linear size~$d$ in the interval $2^{k-1}<d\le 2^k$ are
annealed away.  The time-scale for this process is $\O(e^{\beta k})$, and
differs by a factor of $e^\beta$ from the time-scale associated with the
following stage.  In this limit, the master equation for the coarsening
process can be solved using generating functions.  If $P_k(d)$ is the
probability distribution of lengths $d$ at the beginning of $k$-th stage,
$G_k(z) = \sum_{2^{k-1}<d} P_k(d) z^d$ is the probability generating
function for this distribution, and $H_k(z) \equiv \sum_{2^{k-1}<d\leq 2^k}
P_k(d) z^d$ is the equivalent function for the active domains---those which
will be annealed away during the current stage---the following recursive
relations are obtained~\cite{sollich}
\begin{equation} 
G_{k+1}(z) = 1 + [ G_k(z) - 1 ] \exp\bigl(H_k(z)\bigr).
\end{equation}
Given an initial distribution $P_0(d)$, these equations can be solved
numerically.

\begin{figure}[t]
\begin{center}
\epsfig{file=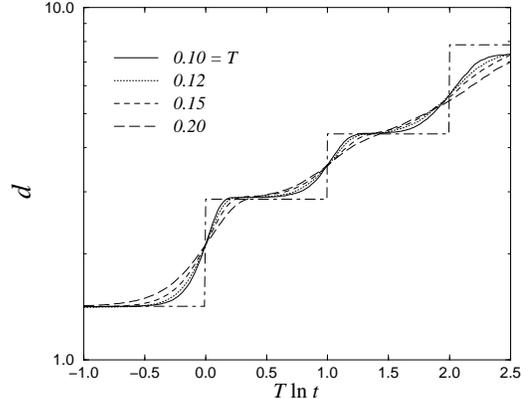,width=\figurewidth}
\end{center}
\caption{Average distance between defects $d$ as a function of rescaled
  time $\nu = T \ln(t)$.  The dot-dashed line corresponds to the $T=0$
  coarsening approximation.}
\label{dTlogt}
\end{figure}

In Fig.~\ref{dTlogt} we show the average linear length of the domains $d(t)
= 1/\sqrt{\varepsilon(t)}$ as a function of $\nu = T \ln t$ from our
simulations.  As $T\to0$, we expect $d(t)$ to become a ``staircase''
function, with the steps extending between integer values of $\nu$.  The
approximate heights of these steps are given by the derivative of the
generating function at $z=1$: $\langle d\rangle_k = \sum_d d P_k(d) =
G_k'(1)$.  These steps are also plotted in the figure (dot-dashed line)
and, as we can see, the agreement is quite good.

\begin{figure}[t]
\begin{center}
\epsfig{file=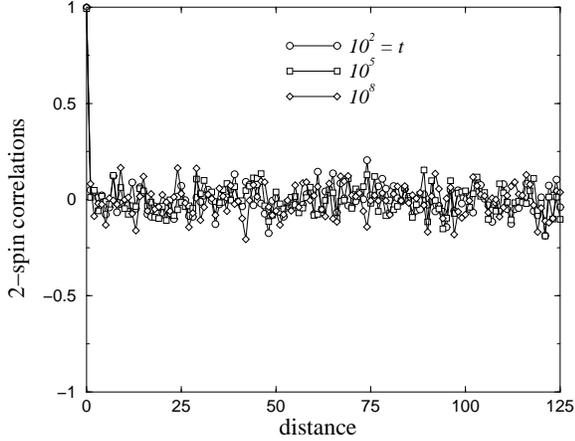,width=\figurewidth}
\end{center}
\caption{Two-spin correlations for various times as a function of distance.}
\label{2spins}
\end{figure}

We now turn to the behaviour of the spin correlation functions as we
approach equilibrium.  In equilibrium, as shown above, the magnetization
and all two-spin correlations vanish at any finite temperature.  Moreover,
at all times following a quench, the average magnetization and all
equal-time 2-spin correlations vanish as well, as shown in
Fig.~\ref{2spins}.  This is a consequence of the three-spin interactions:
regardless of the value of a spin~$i$, the Hamiltonian favours a
neighbouring spin $j$ equally to be up or down, since the only interaction
between the two spins also includes a third spin of unknown orientation.

\begin{figure}[t]
\begin{center}
\epsfig{file=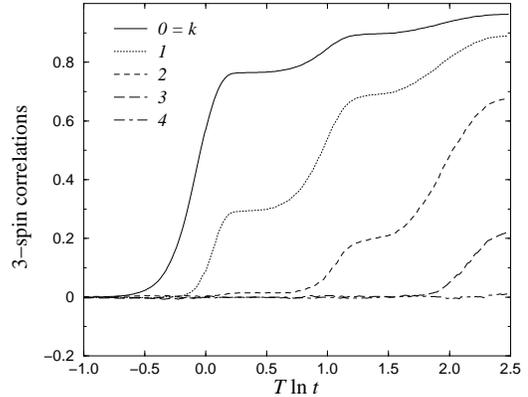,width=\figurewidth}
\end{center}
\caption{Three-spin correlation functions $C_k^{(3)}$ as a function of
  time, for linear sizes $2^{k=0,1,2,3,4}$. We plot the absolute value of
  the correlations.}
\label{3spins}
\end{figure}

In equilibrium, the first non-zero correlations are those for triplets of
spins at the vertices of downward pointing triangles of linear size $2^k$.
This is also the case in the out-of-equilibrium regime following a quench.
In Fig.~\ref{3spins} we show the absolute value of the $3$-spin
correlations $|C_k^{(3)}|$ as a function of time after a quench, for
various values of~$k$.  The behaviour of these curves illustrates the
nature of the stages by which relaxation takes place.  Initially,
correlations at all length-scales are zero.  The system has no barriers to
relaxation of excitations of length-scale~1, so $|C_0^{(3)}|$ starts to
grow immediately after the quench.  At $\nu=T\ln t=0$, $|C_1^{(3)}|$ grows
exponentially fast to a first quasi-stationary value, while all
$C_{k>1}^{(3)}$ remain zero: this first plateau corresponds to partial
equilibration up to length-scales of~2.  At $\nu=1$, the system starts to
relax up to length-scales of~4, with $C_2^{(3)}$ becoming nonzero, while
$C_{k>2}^{(3)} = 0$.  And so forth.

\begin{figure}[t]
\begin{center}
\end{center}
\epsfig{file=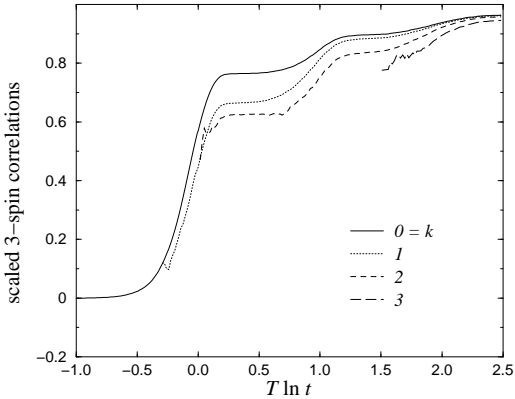,width=\figurewidth}
\caption{Three-spin correlations scaled according to Eq.~\eref{c3k}.}
\label{scaled_3spins}
\end{figure}

Within the plateaus all one-time quantities have roughly stationary values,
and so it is natural to ask to what extent the system is in a
quasi-equilibrium state.  One answer to this question comes from examining
the agreement between our non-equilibrium correlation functions and the
exact relations~\eref{c3k} for the equilibrium correlation function.  In
Fig.~\ref{scaled_3spins} we compare $|C_0^{(3)}|$ with
$|C_k^{(3)}|^{(1/3^k)}$, and find that the out-of-equilibrium correlation
functions still collapse approximately when scaled appropriately, the
collapse getting better as time progresses.  This may indicate that the
later stages can be described by an approximate equilibrium at an
appropriate effective temperature.  We discuss this point further in
Section~\ref{threec}.

\subsection{Two-time quantities}
\label{twotime}
\begin{figure}[t]
\begin{center}
\epsfig{file=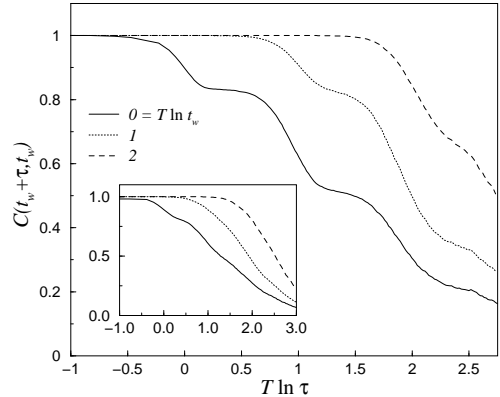,width=\figurewidth}
\end{center}
\caption{Local spin-spin correlation functions as a function of the scaled
  time difference $T \ln \tau$, following a quench to $T=0.12$.  Inset:
  the same for $T=0.20$.}
\label{corrTlogt}
\end{figure}

\begin{figure}[t]
\begin{center}
\epsfig{file=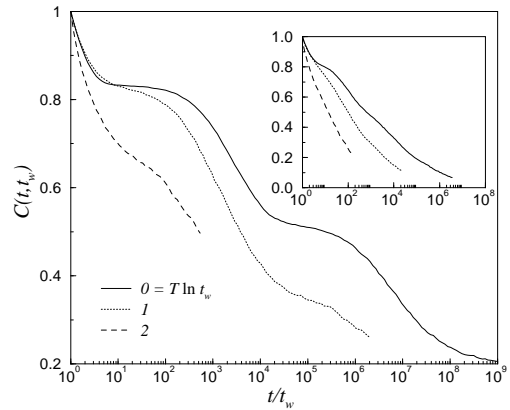,width=\figurewidth}
\end{center}
\caption{Local spin-spin correlation functions as a function of $t/t_w$,
  for $T=0.12$.  Inset: the same for $T=0.20$.}
\label{corrttw}
\end{figure}

We now turn to the behaviour of two-time quantities in the
out-of-equilibrium regime following a quench from high temperature.  We
first consider the local two-time spin autocorrelation function
\begin{equation} 
C(t,t_w) = \frac{1}{N} \sum_{mn} 
           \langle \sigma_{mn}(t) \sigma_{mn}(t_w) \rangle .
\end{equation}
In Fig.~\ref{corrTlogt} we show $C(t,t_w)$ for quenches from $T=\infty$ to
$T=0.12$ for three different values of the weighting time~$t_w$, as a
function of the rescaled time difference $\nu = T \ln \tau$, with
$\tau=t-t_w$.  The three values of $t_w$ used correspond to $\nu = 0$, $1$,
$2$, i.e.,~to the starting points of the first three of the plateaus
discussed in Section~\ref{onetime}.  In Fig.~\ref{corrttw} we present the
same correlations as a function of $t/t_w$.

Since the system does not reach equilibrium at $T=0.12$ on the timescales
simulated, we do not in general expect the correlation functions to be
functions of the time difference $\tau=t-t_w$ only.  In fact, as
Fig.~\ref{corrTlogt} shows, the behaviour of the correlation functions has
a clear dependence on the waiting time.  On the other hand, as we can see
from Fig.~\ref{corrttw}, neither does it obey a simple aging form, scaling
with $t/t_w$, as observed in other cases~\cite{review}.  This is
because the length-scale associated with the relaxation of the spins does
not grow as a simple power of time (see Figs.~\ref{dTlogt}
and~\ref{3spins}).  In general, if equilibration is associated with the
growth of a length-scale~$l(t)$, then the two-time correlation functions
should scale with $l(t)/l(t_w)$~\cite{review}.  Behaviour of this kind was
found in the autocorrelation functions for the ACIC~\cite{felix}, where the
appropriate length-scale is the average distance between upward-pointing
spins.  In our model, however, rescaling the spin correlation functions by
the distance between defects fails, and we have not been able to find a
suitable length-scale such that the scaling with $l(t)/l(t_w)$ holds.

Another observable of interest is the response function $\chi(t,t_w)$,
which measures the response of the spins at time $t$ to a small field
applied at time $t_w$.  Comparison of such a response function with the
two-time correlations studied above can reveal violations of the
fluctuation--dissipation theorem (FDT) which are expected in systems
displaying aging.  In general, in order to obtain the response function
corresponding to the {\em local\/} correlation function calculated above,
we would need to apply a field to only a single spin on the lattice, or
equivalently we could apply a random field and measure the staggered
response~\cite{barrat}.  Both of these approaches present significant
numerical challenges, and require a substantial investment of CPU time in
simulation to extract clean results.  For our model, however, this turns
out to be unnecessary because, as mentioned in Section~\ref{onetime}, all
off-diagonal spin-spin correlations vanish, implying that two-time
autocorrelations of the magnetization $m \equiv N^{-1} \sum_{mn}
\sigma_{mn}$ are proportional to the local spin correlations thus:
\begin{equation} 
\langle m(t) m(t') \rangle = {1\over N} C(t,t').
\end{equation}  
This means that we need only to measure the response to a uniform field.

\begin{figure}[t]
\begin{center}
\epsfig{file=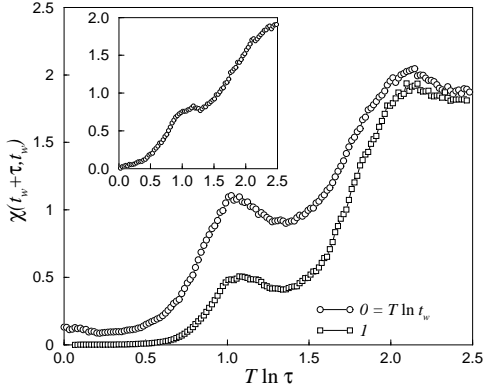,width=\figurewidth}
\end{center}
\caption{Integrated response function vs.\ rescaled time difference, for
  $T=0.12$ and $T \ln t_w = 0$ and $1$.  Inset: the same for $T=0.20$.}
\label{res}
\end{figure}

We have performed simulations in which the Hamiltonian was perturbed with a
small uniform magnetic field applied at a time $t_w$ following a quench to
finite temperature: $\Delta H(t) = h_0\theta(t-t_w)\sum_{mn}\sigma_{mn}$.
We measured the integrated response (or relaxation function)
\begin{equation} 
\chi(t,t_w) = \frac{1}{N h_0} \sum_{nm} 
              \langle \sigma_{nm}(t) \rangle_h,
\end{equation}
which is equivalent to measuring the change in magnetization per spin of
the system, divided by the strength of the applied field.  We have
experimented with a variety of different values for the field strength
$h_0$; the results presented here are for $h_0=0.02$, which we find to be
well within the linear regime.  In Fig.~\ref{res} we show the measured
response for $T=0.12$ and $\nu_w = T \ln t_w = 0$ and~1.  Notice that the
response is not in general a monotonically increasing function of time, but
instead shows cusps at times corresponding to the jumps between plateaus.
For higher temperatures the cusps are less pronounced and eventually
disappear (see inset).

A similar non-monotone behaviour is seen in the response function for the
defects in our model, and can be understood physically by the following
argument.  The energy density $\varepsilon$ plays a role for the defects
equivalent to that played for the spins by the magnetization~$m$, and the
equivalent of a small applied magnetic field $\delta h$ is a small change
in temperature $\delta T = - T^2 \, \delta h$.  Consider then the behaviour
of the energy density shown in Fig.~\ref{et}.  At low temperatures we
expect $\varepsilon$ to follow a staircase function, with steps extending
between integer values of~$\nu$.  When the field is applied (i.e.,~when the
temperature is changed), the perturbed energy density is another staircase
function, with each step having a slightly higher (lower) value for
negative (positive) field, and lasting a longer (shorter) time, the steps
now extending between integer values of $\nu (1- T\, \delta h)$.  The
corresponding integrated response is given by the difference between the
perturbed and unperturbed energy densities, divided by the field.  Within
each plateau this difference is mostly small and roughly constant, but a
sharp change is seen at the boundaries between steps.  The unperturbed
($T\to0$) density jumps to the next plateau when $\nu$ is an integer, but
the perturbed one only does so when $\nu (1- T\, \delta h)$ is an integer,
and thus the absolute value of the response is large during the short
interval between these two events.  Once the perturbed density jumps to the
next plateau, the response becomes small again.  This increase and decrease
at integer $\nu$ is responsible for the humps seen in the integrated
response.  The argument generalizes to perturbations with a staggered
random field.

The response function for the ACIC was studied in Ref.~\onlinecite{felix},
where it was found to be a monotonically increasing function of time.
However, as mentioned above, the defect representation of our model behaves
very similarly to the ACIC, and we therefore conjecture that non-monotone
behaviour will be seen in the ACIC also at sufficient low temperatures, for
precisely the reasons given above.

\subsection{Configurational entropy and fluctuation--dissipation relations}
\label{threec}
A configuration of our model is a local energy minimum if and only if no
two defects occupy adjacent sites on the lattice.  In the context of
glasses, such states are known as inherent structures~\cite{is}.  A nice
feature of the model is that the set of inherent structures is isomorphic
to the set of allowed states of Baxter's hard-hexagon model~\cite{baxter},
allowing the distribution of inherent structures to be calculated exactly.

The grand-partition function of the hard-hexagon model on a lattice of $N$
sites is
\begin{equation} 
Z_N(z) = \sum_{M=0}^N \, g(M,N) \, z^M ,
\label{z1}
\end{equation}
where $z$ is the fugacity and $g(M,N)$ is the number of ways of placing $M$
non-adjacent particles on the triangular lattice.  In our case,
$M=N\varepsilon$ corresponds to the total energy of our model, so
$g_N(\varepsilon) \equiv g(N \varepsilon,N)$ gives the density of states
for inherent structures with energy~$\varepsilon$.  For large $N$, $g(n,N)$
is exponential in $N$, so that $g_N(\varepsilon) =
e^{NS_c(\varepsilon)}$, where $S_c(\varepsilon)$ is the
configurational entropy of the model, i.e.,~the entropy density of
metastable states.  In this case, Eq.~\eref{z1} becomes
\begin{equation}
Z_N(z) = \int d\varepsilon \> \exp \bigl[ N
        \bigl( \varepsilon \ln z + S_c(\varepsilon) \bigr) \bigr].
\end{equation}
For large $N$ the integral is dominated by the saddle point in the
exponent, and we obtain
\begin{eqnarray}
\label{kappa1}
\varepsilon(z) &=& \frac{\partial\ln\kappa}{\partial\ln z},\\
\label{kappa2}
S_c[\varepsilon(z)] &=& \ln \kappa(z) - \varepsilon(z)\ln z,
\end{eqnarray}
where $\kappa(z) \equiv \lim_{N \to \infty} [Z_N(z)]^{1/N}$ is the
partition function per site of the hard-hexagon model, which is known
exactly in the thermodynamic limit~\cite{baxter}.  Between them,
Eqs.~\eref{kappa1} and~\eref{kappa2} determine $S_c(\varepsilon)$
parametrically.

At low defect densities, $S_c(\varepsilon)$ reduces to
\begin{equation}
S_c(\varepsilon) = - \varepsilon \ln \varepsilon + \O(\varepsilon),
\label{eloge}
\end{equation}
which is the general form for a low concentration of non-interacting point
defects in an ordered structure~\cite{stilli}.  The configurational
entropies of the disordered Ising chain, constrained Ising chains, and the
Backgammon model all have this asymptotic form~\cite{felix,bm}.  Note that
$S_c$ has infinite slope with respect to $\varepsilon$ at
$\varepsilon=0$ where it vanishes, which is consistent with the fact that
none of these models has a finite temperature phase transition.  However,
since $S_c'$ increases logarithmically as $\varepsilon$ is decreased, an
extrapolation of $S_c'(0)$ from the behaviour at finite $\varepsilon$
would wrongly suggest a finite slope and therefore a finite Kauzmann
temperature.  It has been argued that this mechanism would rule out an
ideal-glass phase transition in materials composed of limited size
molecules and conventional molecular interactions~\cite{stilli}.

Having calculated correlation and response functions for the spins in our
model (Section~\ref{twotime}), we can use our results to study the relation
between fluctuations and responses, which in some other systems is related
to the configurational entropy.  The natural way to do this is by means of
a parametric plot of $\chi(t_w+\tau,t_w)$ versus $C(t_w+\tau,t_w)$ for
fixed $t_w$~\cite{cuku}.  For a system in equilibrium at temperature~$T$,
such a plot would be linear with slope $-1/T$, in accordance with the
fluctuation--dissipation theorem.  In a glassy system on the other hand,
the FDT is normally violated in the out-of-equilibrium low-temperature
regime as $t_w \to \infty$, and this violation encodes important
information about the system's dynamical behaviour (see
Ref.~\onlinecite{review} for a review).  The classic example is the
mean-field $p$-spin spin-glass~\cite{pspin}, for which the FDT plot is
piecewise linear: for large values of $C$, corresponding to relaxation of
fast degrees of freedom, the plot has a slope of $-1/T$ as it would in
equilibrium, but for smaller $C$, corresponding to the relaxation of slow
modes, it has slope $-1/T_{\rm eff}$, where $T_{\rm eff}>T$ is interpreted
as a effective temperature for these modes~\cite{cukupe}.  Furthermore, in
the $p$-spin model $T_{\rm eff}$ is numerically equal to the inverse of the
slope of the configurational entropy at the asymptotic energy
density~\cite{mona}.  Similar behaviour has also been observed in more
realistic model glass formers~\cite{kob,lapr}.

\begin{figure}[t]
\begin{center}
\epsfig{file=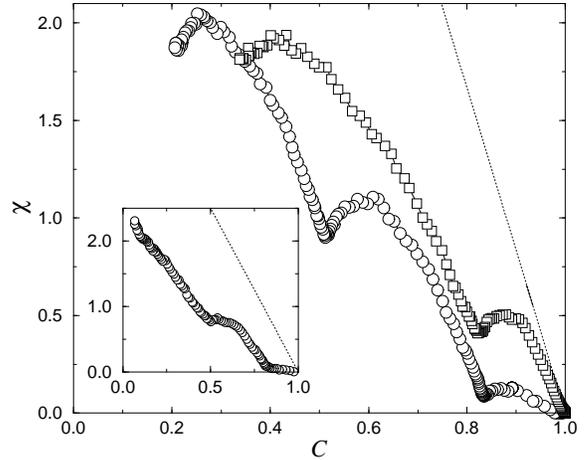,width=\figurewidth}
\end{center}
\caption{Parametric plot of integrated response vs. the two-time
  autocorrelation function of the spins for $T=0.12$ and $T \ln t_w = 0$
  and $1$.  The dotted line corresponds to slope $-1/T$.  In the inset we
  show the same for $T=0.20$ and $T \ln t_w = 0$.}
\label{CR}
\end{figure}

At sufficiently low temperatures our model is clearly out of
equilibrium, and although it does not reach the long-time asymptotic
regime corresponding to approach to equilibrium in our simulations, it
is possible that the FDT plots can still provide information on the
relaxation process, and maybe that there exist effective temperatures
associated with the different time-scales in the problem
\cite{cukupe}.  In Fig.~\ref{CR} we show the FDT plot for the spin
response and correlation functions.  The unusual non-monotonic shape
is a consequence of the non-monotonicity of the response function.

The curves should be ``read'' from right to left in the plot, and are
composed of a sequence of segments, each associated with one of the
plateaus.  The starting point of each segment and the part of the curve in
which the response is increasing correspond to the time the system spends
within the relevant plateau.  The maximum and the downward portion of the
segment correspond to the transition to the next plateau.  The first part
of each segment has a shape similar to that found in
Refs.~\onlinecite{gluck} and~\onlinecite{crising} for the one-dimensional
Ising model following a quench into the temperature scaling region: the FDT
curve there has slope $-1/T$ at $C=1$, and bends continuously with
decreasing $C$ to reach a slope of $-1/2T$ at $C=0$.  (Note that this
behaviour is very different from that seen in domain growth in higher
dimensions~\cite{barrat}.)  In our model, each of the segments of the FDT
plot starts with slope approximately $-1/T$ (indicated by the dotted line
in the figure), but the subsequent shape of the curve varies from one
segment to another.  The similarity within the plateaus to the behaviour of
the one-dimensional Ising model is consistent with what was found for the
statics of the model in Section~\ref{statics}.

An important open question is whether the configurational entropy of
Eqs.~\eref{kappa1} and~\eref{kappa2} plays any role in the
out-of-equilibrium dynamics.  A plausible explanation for the shape of the
FDT curves of Fig.~\ref{CR} is that the initial part of each segment
corresponds to quasi-stationary thermal excitations of fast modes, while
the latter part corresponds to slower large-scale rearrangements arising
from occasional jumps between local minima before the transition to the
next plateau takes place.  If the slope of the latter part of each segment
corresponds to an effective temperature for that segment, we might expect
these temperatures to be related to the rate of change of
$S_c(\varepsilon)$ at the energy density of the associated plateau.  It
is noteworthy that the final slopes of each segment in Fig.~\ref{CR} are
roughly equal to $1/S_c'(\varepsilon_k)$, with values approximately
$\frac12$, $\frac13$, and $\frac14$ for the first three plateaus.  This
observation is however speculative and furthermore contrasts with the
conclusions of Ref.~\onlinecite{felix} for the ACIC, so it deserves further
investigation.

\section{Conclusions}
\label{concs}
In this paper we have studied the low temperature behaviour of the glassy
two-dimensional spin model with uniform ferromagnetic short-range
three-spin interactions introduced by Newman and Moore.  The model has a
dual description in terms either of interacting spins or of free defects,
the mapping between the spin and defect representations being one-to-one.
This allows us to compute exactly all equilibrium correlation functions,
both for spins and defects.  We have also shown that the model displays no
static phase transition at finite temperature.

Despite the simplicity of its statics, the model's low-temperature
single-spin-flip dynamics is highly non-trivial.  A spin flip in the model
corresponds to the flipping of three neighbouring defects, which implies
that at low temperatures the dynamics of the defects is constrained: a
defect can only be flipped if another neighbouring defect is present.  This
in turn implies that the relaxation of isolated defects is an activated
process, and the size of the corresponding energy barriers are found to
grow logarithmically with the distance between defects.  The dual
representation of the model in terms of strongly interacting spins with a
simple dynamics, and of free defects subject to kinetic constraints, is an
explicit realization of the hierarchical constrained dynamics scenario of
Palmer~\etal~\cite{palmer}.

We have studied in detail the dynamics of the model after a quench to low
temperatures.  The presence of logarithmically growing barriers leads to
non-Arrhenius relaxation, the equilibration time being of the exponential
inverse temperature squared form $t_{\rm eq} \sim \exp(1/[2 T^2 \ln 2])$.
One-time quantities such as internal energy display ``plateaus'' in their
equilibration profiles, which correspond to the trapping of the system in
local energy minima.  Each plateau is associated with a specific stage in
the dynamics, the $k^{\rm th}$ plateau corresponding to the partial
equilibration of length-scales up to $2^k$.  The behaviour of observables
related to the defects is strikingly similar to that seen in the
asymmetrically constrained Ising chain.  For example, in the $T\to0$ limit
the average distance between defects can be well approximated using the
analytic methods of Sollich and Evans~\cite{sollich} which yield exact
results for the asymmetrically constrained model.

We have also studied two-time quantities for the model, like spin
autocorrelations and response functions.  At low temperatures, the
response functions have the unusual property of being non-monotonic:
they display humps at exactly those times at which the system jumps
between plateaus.  This behaviour has also been observed in other
models at times well within the activated regime, such as the
constrained Ising chains and the Backgammon model in one
dimension~\cite{felix}, models for two-dimensional
froths~\cite{lexie}, and vibrated granular media~\cite{nicodemi}.  An
important open question is whether this is a generic feature of the
out-of-equilibrium dynamics of activated processes.

\section*{Acknowledgements}
The authors would like to thank Jean-Philippe Bouchaud, Paul Goldbart,
Jorge Kurchan, Felix Ritort, David Sherrington, and Peter Sollich for
useful discussions.  JPG wishes to thank the Santa Fe Institute and
MEJN the Subdepartment of Theoretical Physics, University of Oxford,
for their kind hospitality while this work was carried out.  This work
was funded in part by EC Grant No.\ ARG/B7-3011/94/27 and EPSRC Grant
No.\ GR/M04426.

\end{document}